# Dalton's law of partial optical thermodynamic pressures in highly multimoded nonlinear photonic systems

HUIZHONG REN,[1] GEORGIOS G. PYRIALAKOS,[1,3] FAN O. WU,[3] NIKOLAOS K. EFREMIDIS,[4] MERCEDEH KHAJAVIKHAN,[1,2] AND DEMETRIOS N. CHRISTODOULIDES[1,2,*]

[1] Ming Hsieh Department of Electrical and Computer Engineering, University of Southern California, Los Angeles, CA 90089, USA
[2] Department of Physics and Astronomy, University of Southern California, Los Angeles, CA 90089, USA
[3] CREOL, The College of Optics and Photonics, University of Central Florida, Orlando, Florida 32816-2700, USA
[4] Department of Mathematics and Applied Mathematics, University of Crete, Heraklion, Crete 70013, Greece
*Corresponding author: demetri@usc.edu

**We show that in highly multimoded nonlinear photonic systems, the optical thermodynamic pressures emerging from different species of the optical field obey Dalton's law of partial pressures. In multimode settings, the optical thermodynamic pressure is defined as the conjugate to the extensive variable associated with the system's total number of modes and is directly related to the actual electrodynamic radiation forces exerted at the physical boundaries of the system. Here, we extend this notion to photonic configuration supporting different species of the optical field. Under thermal equilibrium conditions, we formally derive an equation that establishes a direct link between the partial thermodynamic pressures and the electrodynamic radiation pressures exerted by each polarization species. Our theoretical framework provides a straightforward approach for quantifying the total radiation pressures through the system's thermodynamics variables without invoking the Maxwell stress tensor formalism. In essence, we show that the total electrodynamic pressure in such arrangements can be obtained in an effortless manner from initial excitation conditions**, **thus avoiding time-consuming simulations of the utterly complex multimode dynamics. To illustrate the validity of our results, we carry out numerical simulations in multimoded nonlinear optical structures supporting two polarization species and demonstrate excellent agreement with the Maxwell stress tensor method.**

Pressure effects arising from electrodynamic forces have so far found applications in many and diverse fields, such as laser cooling, optical tweezers and in manipulating microscopic objects, to mention a few [1]. Recently, advances in optomechanics have opened up altogether new possibilities through the manipulation of photon-phonon coupling processes [2]. Along similar lines, radiation pressure phenomena have also been systematically investigated in waveguide and cavity arrangements. In waveguide systems, bonding and anti-bonding radiation forces between two single-mode evanescently coupled channels have been theoretically predicted and experimentally observed on integrated optical platforms [3, 4]. In addition, radiation pressure effects have been explored in high-contrast waveguides for engineering their stimulated Brillouin scattering performance [5].

Thus far, the theoretical analysis of electromagnetic forces has predominantly relied on two approaches, which have been demonstrated to be equivalent [6]: (a) the Maxwell stress tensor formalism where the force can be obtained by integrating the corresponding tensor element over a closed surface and (b) an energy-based method from which one can extract the force through the variation of the propagation constant/eigenvalue [3]. These two methodologies require that either the associated electromagnetic vectorial field distribution or the variation of the eigenvalue must be known a priori. In this respect, calculating the radiation forces using the aforementioned methods is straightforward in few-mode settings. However, in highly multimoded nonlinear optical arrangements, the evaluation of radiation forces becomes a tedious task, given that one has to first extract the vectorial electromagnetic field distributions before even attempting to solve the convoluted evolution equations associated with the nonlinear dynamics of possibly thousands of modes.

To tackle this issue, based on the newly developed optical thermodynamic theory [7], a new methodology has been put forward that enables the evaluation of electrodynamic radiation forces in equilibrated multimoded nonlinear structures [8, 9]. This approach demands the presence of an intensive quantity, the so-called "optical thermodynamic pressure" $\hat{p}$ that so happens to be conjugate to the extensive variable associated with the system's total number of modes $M$ [7, 10]. This quantity complements the other two primary intensive variables, the optical temperature $T$ and the optical chemical potential $\mu$. At a fundamental level, these variables arise from a statistical mechanics treatment of the nonlinear dynamics, bearing no direct correspondence to their physical counterparts (e.g. the optical temperature is unrelated to the physical temperature of the environment the photonic system is embedded in). Nonetheless, they have a tangible physical manifestation within a photonics context as they represent actual thermodynamic forces that dictate the statistical exchange of energy or power between thermalizing optical subsystems [7]. Quite recently it was found that [9], unlike $T$ and $\mu$, $\hat{p}$ can be directly related to the actual electromagnetic radiation pressure $p^{EM}$, owing to a direct correspondence between $M$ and a fiber's radius $a$. Analogous to the ideal gas law, this led to a novel approach for calculating electromagnetic radiation pressure effects using statistical thermodynamic quantities, which, in turn, can be extracted through the initial excitation conditions [11].

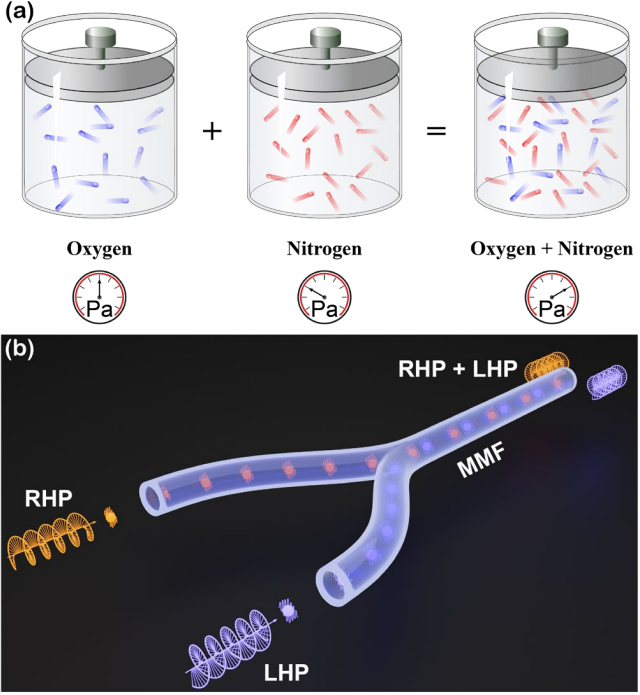

Fig. 1. (a) Dalton's law of partial pressures in a mixture of non-reacting gases. The total pressure of the gas mixture is given by the sum of the partial pressures. (b) A nonlinear multimode fiber (MMF) allows two "non-reacting" circular polarizations to reach thermal equilibrium, where the optical version of Dalton's law can be studied.

In 1802, John Dalton formulated the law of partial pressures. By assuming ideal gas conditions, he showed that the total pressure in a mixture of non-reacting gases is equal to the sum of partial pressures of these species (Fig.1 (a)). At this point, a question arises as to whether a Dalton's law of partial optical thermodynamic pressures can be established in multimode nonlinear photonic arrangements, and if so, to what extent it applies to the actual electrodynamic pressures.

In this Letter, we demonstrate that in highly multimoded nonlinear photonic systems, the optical thermodynamic pressures emerging from different species of the optical field obey Dalton's law of partial pressures. We find that in a mixture of "non-reacting photon gases" (like two polarizations), the total thermodynamic pressure is equal to the sum of the partial pressures associated with each individual component. In the context of optical thermodynamics, a "photon gas" can attain thermal equilibrium, under the chaotic and ergodic action of nonlinearity when propagating in a multimode optical system [7]. As previously shown, upon equilibration, the photon gas statistics (power allocation per mode) obey a Rayleigh-Jeans distribution that is characterized by an optical temperature and a chemical potential. In principle, a mixture of two different "non-reacting photon gases" can be established when the same nonlinear fiber involves two polarizations, as long as they do not exchange optical power (Fig.1 (b)). This is possible through the use of the two circular polarizations in multimode cylindrical fibers or of the two linear polarizations in highly birefringent waveguide systems. In such settings, the two species interact only through cross-phase modulation (XPM) effects while the power in each species is conserved [12]. At the same time, XPM leads to an exchange of longitudinal momentum, thus allowing the two species to reach thermalization at the same optical temperature, as expected by the second law of thermodynamics. We here employ the tools of optical thermodynamics to formally derive an equation that unifies the total optical thermodynamic pressure $\hat{p}$ with the electrodynamic pressure $p^{EM}$ exerted at the boundaries of the structure.

We begin our analysis by considering two "non-reacting" polarizations propagating in a nonlinear multimode optical waveguide. Each component supports a finite number of $M$ bound modes and propagates along the axial direction $z$. The eigenmode $|\psi_i\rangle$ and its corresponding propagation constant $\epsilon_i$ satisfy the eigenvalue problem $\hat{H}_L|\psi_i\rangle = \epsilon_i|\psi_i\rangle$, where $\hat{H}_L$ is the linear Hamiltonian operator of the optical system. Without any loss of generality, we assume that the eigenvalues are arranged in ascending order, according to $\epsilon_1 \geq \epsilon_2 \geq \cdots \geq \epsilon_M$, where $\epsilon_1$ is the eigenvalue of the ground state while $\epsilon_M$ corresponds to the eigenvalue of the highest-order mode supported in this system. Similar to a mixture of inert gases where the number of particles of individual constituents remains invariant, in this optical arrangement, only exchange of optical "internal energy" is allowed, while the power of each component remains constant. In this respect, one can identify the following conservation laws:

$$\mathcal{P}_1 = \sum_{i=1}^{M} n_{1,i} \quad \mathcal{P}_2 = \sum_{i=1}^{M} n_{2,i}, \quad (1)$$

$$U = U_1 + U_2 = -\sum_{i=1}^{M} \epsilon_i n_{1,i} - \sum_{i=1}^{M} \epsilon_i n_{2,i}, \quad (2)$$

where $\mathcal{P}_k$ stands for the power in each species and $U$ is the total "internal energy" of the system (longitudinal momentum) [13]. In the above expressions, $n_{k,i}$ represents a quantity that is proportional to the average power $|c_{k,i}|^2 = \mathcal{P}_0 n_{k,i}$ conveyed by mode $|\psi_i\rangle$ of component $k = 1,2$ where the arbitrary level $\mathcal{P}_0$ denotes the power of each discrete power-packet. At thermal equilibrium, the power distribution in each polarization $n_{k,i}$ can be determined by maximizing the system's Boltzmann entropy under the constraints provided by the three invariants of Eqs. (1-2):

$$S = S_1 + S_2 = \sum_{i=1}^{M} \ln n_{1,i} + \sum_{i=1}^{M} \ln n_{2,i} \quad (3)$$

where $S_k$ denotes the entropy of each individual constituent. Note that in obtaining Eq. (3), we made use of the fact that entropy is extensive. To maximize the total entropy, we use Lagrange multipliers [7, 10, 14]. The corresponding Lagrange function $\mathcal{L}$ is given by

$$\mathcal{L} = S + \sum_k \alpha_k \left( \sum_{i=1}^{M} n_{k,i} - \mathcal{P}_k \right) + \beta \left( \sum_k \sum_{i=1}^{M} \epsilon_i n_{k,i} + U \right),$$

where $\alpha_k$ and $\beta$ denote the Lagrange multipliers. Extremization of entropy demands that

$$\frac{\partial \mathcal{L}}{\partial n_{k,i}} = 0$$

from which one can derive

$$n_{k,i} = -\frac{1}{\alpha_k + \beta \epsilon_i}.$$

By adopting $\alpha_k = \mu_k T^{-1}, \beta = T^{-1}$ [7, 14, 15], we obtain the Rayleigh-Jeans distribution for each of the two photon species

$$n_{k,i} = -\frac{T}{\epsilon_i + \mu_k}, \quad (4)$$

where $T$ is the common optical temperature and $\mu_k$ denotes the corresponding chemical potential. In this respect, Eq. (4) indicates that all constituents will eventually thermalize into a Rayleigh-Jeans distribution, ultimately reaching a common optical temperature $T$ but different chemical potentials $\mu_k$. This is in direct analogy to two non-reacting gases coexisting in a shared container (Fig. 1(a)), where they can exchange energy without chemically reacting, ultimately reaching a common temperature. Here, it is worth noting that the optical temperature and chemical potentials can be determined uniquely from the initial excitation conditions [11]. From the Rayleigh-Jeans distribution, one can formally derive the following two global equations of state [7]:

$$U_k - \mu_k \mathcal{P}_k = MT \quad (5)$$

which relate the intensive variables $\mu_k$ and $T$ to the extensive quantities $U_k, \mathcal{P}_k$ and $M$ for each individual component.

As shown in previous studies, the entropy $S_k$ of each constituent in such a microcanonical optical system can be expressed as a function of three extensive variables, i.e., $S_k = S_k(U_k, \mathcal{P}_k, M)$. In addition, the extensivity of the entropy demands that $\sum_k S_k(\lambda U_k, \lambda \mathcal{P}_k, \lambda M) = \lambda \sum_k S_k(U_k, \mathcal{P}_k, M) = \lambda S$. From here one can identify the following conjugate intensive variables: the optical temperature $T$, the chemical potentials $\mu_k$ and the optical thermodynamic pressures $\hat{p}_k$, via

$$\frac{1}{T} = \frac{\partial S_k}{\partial U_k}, \quad \frac{\mu_k}{T} = \frac{\partial S_k}{\partial \mathcal{P}_k}, \quad \frac{\hat{p}_k}{T} = \frac{\partial S_k}{\partial M}. \quad (6)$$

These intensive quantities act in every respect as thermodynamic forces [7]. In this regard, one can formally derive the Euler equation

$$\sum_k (U_k + \hat{p}_k M - \mu_k \mathcal{P}_k) = T \sum_k S_k = TS \quad (7)$$

By substituting Eq. (5) into Eq. (7), we obtain

$$\hat{p}_T = \hat{p}_1 + \hat{p}_2 = T \left( \frac{S_1}{M} - 1 + \frac{S_2}{M} - 1 \right) = T \left( \frac{S}{M} - 2 \right) \quad (8)$$

Equation (8) implies that, in a mixture of non-reacting photon gases, the exerted total optical thermodynamic pressure is equal to the sum of the two partial pressures associated with each individual component. Equation (8), a central result in our study, formally corresponds to a Dalton's law for partial optical thermodynamic pressures. Next, we will demonstrate that the total optical thermodynamic pressure $\hat{p}_T$ is directly related to the actual total electromagnetic pressure $p_T^{EM}$.

Recently, the physical nature of the so-called optical thermodynamic pressure $\hat{p}$ has been investigated in optical nonlinear multimode arrangements when a single species is involved [9]. In this respect, it was found that this optical thermodynamic quantity is directly linked to the total radiation pressure $p^{EM}$, once a universal entropic component is taken into account. In general, at thermal equilibrium, the quantities $\hat{p}$ and $p^{EM}$ are formally related via [9]

$$\hat{p} = Q \cdot p^{EM} + T \ln \left( -\frac{T}{\epsilon_M + \mu} \right) - T \quad (9)$$

where $\epsilon_M$ represents the eigenvalue of the highest-order mode while the prefactor $Q$ is completely determined by the system itself. In the same vein, it is straightforward to show that in a composite system involving $k$ "non-reacting" optical components, the following relation holds

$$\hat{p}_T = Q \sum_k p_k^{EM} + \sum_k \left[ T \ln \left( -\frac{T}{\epsilon_M + \mu_k} \right) - T \right] \quad (10)$$

where $p_k^{EM}$ is the electromagnetic radiation pressure exerted by constituent $k$. In this regard, by substituting Eq. (8) into Eq. (10), one finds

$$p_T^{EM} = \frac{T}{Q} \left[ \frac{S}{M} - \sum_k \ln \left( -\frac{T}{\epsilon_M + \mu_k} \right) \right] \quad (11)$$

where $p_T^{EM} = \sum_k p_k^{EM}$ denotes the total electromagnetic pressure exerted at the boundaries of a multimode waveguide at thermal equilibrium. Equation (11) implies that the actual radiation pressure in such multi-species system can be directly obtained through thermodynamic quantities in an effortless manner, without the need to delve into tedious Maxwell stress tensor calculations that require full knowledge of the vectorial electromagnetic fields [16]. Here, if one knows the linear spectrum of propagation constants $\epsilon_i$, as well as the initial excitation conditions that specify $\mathcal{P}_1, \mathcal{P}_2, U$, then it is possible to calculate $S, T, \mu_k$ and therefore $p_T^{EM}$ from Eq. (11).

To verify the validity of our theoretical results, we investigate electromagnetic radiation pressure effects in a scenario where two circularly polarized beams are launched into a nonlinear multimode weakly guiding step-index silica fiber. This fiber supports $M = 234$ LP modes in each polarization and is excited with 2000 kW of power. The total electric field propagating in the fiber is expressed as $\vec{E} = u|R\rangle + v|L\rangle$, where $|R\rangle$ and $|L\rangle$ denote the right- and left-hand circular polarization, respectively and $u, v$ are the field envelope amplitudes. The evolution equations are included in Supplementary. Note that the quantities $\mathcal{P}_u$, $\mathcal{P}_v$ and $U = U_u + U_v$ remain invariant during propagation. At the input, $\mathcal{P}_u(0) = 800$ kW, $\mathcal{P}_v(0) = 1200$ kW, $U_u(0) = -1.14 \times 10^5$ kW/cm and $U_v(0) = -4.16 \times 10^5$ kW/cm.

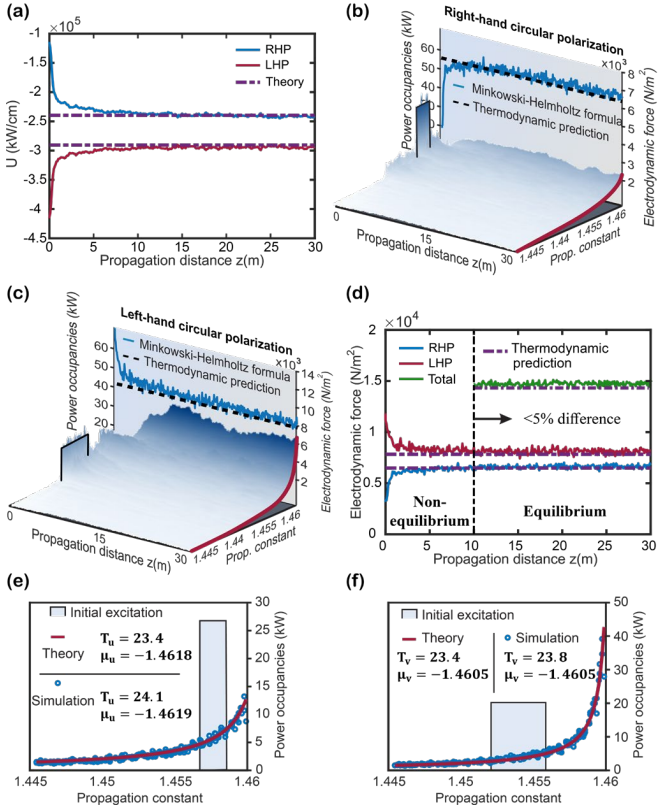

**Fig. 2.** (a) The two circular polarizations exchange energy $\Delta U$ until thermal equilibrium is attained. (b, c) Evolution of the modal occupancies and electrodynamic pressures as a function of the distance associated with the RHP and LHP. (d) The total electrodynamic pressure (green line) obtained from Eqs. (11) and (13). The two results are in excellent agreement, confirming the validity of the optical Dalton's law of Eq. (11). (e,f) Two mode groups are initially excited (shaded areas) for the RHP and LHP photon gases. The RJ distributions along with the optical temperature and chemical potentials are provided, as obtained from simulations and theory. The dotted lines in (a-d) correspond to the theoretically anticipated values.

Our numerical simulations show that, during propagation, these two components progressively exchange their internal energies, until they converge to the Rayleigh-Jeans state predicted by Eq. (4). At thermal equilibrium, they attain the same optical temperature $T = 23.4$ while the chemical potentials settle to $\mu_u = -1.4618$ and $\mu_v = -1.4605$. The electromagnetic pressure at each point $z$ is monitored by performing a spatial average of the local radiation forces per unit area along the circumference of the fiber, i.e., $p^{EM} = (2\pi)^{-1} \int_0^{2\pi} f(\theta) d\theta$, where $f(\theta)$ can be obtained from the Minkowski-Helmholtz formula [6, 16]

$$f(\theta) = \frac{1}{4} \epsilon_0 (n_1^2 - n_2^2) |E(r=a)|^2 \quad (13)$$

In Eq. (13), $\epsilon_0$ is the vacuum permittivity and $E(r=a)$ represents the electric field at the boundary. As Fig. (2) shows, during thermalization, the electromagnetic pressure $p_T^{EM}$ gradually relaxes into its equilibrium value. Here, we confirm that this equilibrium value can be evaluated directly using Eq. (11), where $Q = 4\pi c \mathcal{P}_0^{-1}/[k_0(n_1^2 - n_2^2)]$ and $\epsilon_M \approx k_0 n_2$ with $k_0 = \omega/c$. The entropy $S$ in Eq. (11) is computed from Eqs. (3,4). The simulation results show that these two methods are in excellent agreement with each other. It is worth noting that, in contrast to conventional approaches where a full simulation of the nonlinear propagation dynamics is required for each polarization component, to extract the electromagnetic fields information, Eq. (11) only involves thermodynamic quantities which can be completely determined from the input beam profile.

**Funding.** Air Force Office of Scientific Research (MURI: FA9550-20-1-0322), Office of Naval Research (MURI: N00014-20-1-2789), Army Research Office (W911NF-23-1-0312), National Science Foundation (CCF-2320937), MPS Simons collaboration (Simons grant 733682), Israel Ministry of Defense (IMOD: 4441279927).

**Disclosures.** The authors declare no conflicts of interest.

**Data availability.** Data underlying the results presented in this Letter may be obtained from the authors upon reasonable request.

**Supplemental document.** See Supplementary for supporting content.